\documentstyle[aps,preprint]{revtex}
\input psfig      
\begin{document}      
\draft     

\title{Crossover from Regular to Chaotic Behavior in the  
Conductance of Periodic Quantum Chains} 
 
\author{J. Cserti}
\address{ Department of Physics and Astronomy, The University of Edinburgh,
Edinburgh, EH9 3JZ, Scotland, United Kingdom}
\author{G. Sz\'alka and G. Vattay}  
\address{E\"otv\"os University, Solid State Physics Department, 
H-1088 Budapest, M\'uzeum krt.  6-8, Hungary}   
\date{\today}   
\maketitle 
\begin{abstract} 
The conductance of a waveguide  containing finite number of periodically 
placed  identical  point-like  impurities is  investigated.  It has been 
calculated as a function of both the impurity strength and the number of 
impurities  using the  Landauer-B\"uttiker  formula.  In the case of few 
impurities  the  conductance is  proportional  to the number of the open 
channels $N$ of the empty  waveguide and shows a regular  staircase like 
behavior with step  heights  $\approx  2e^2/h$.  For  large  number of 
impurities the influence of the band structure of the infinite  periodic 
chain can be observed and the conductance is approximately the number of 
energy bands (smaller than $N$) times the universal  constant  $2e^2/h$. 
This lower  value is reached  exponentially  with  increasing  number of 
impurities.  As the  strength of the  impurity is  increased  the system 
passes from integrable to  quantum-chaotic.  The conductance, 
in units of $2e^2/h$, changes from $N$ corresponding to the empty  
waveguide to $\sim N/2 $  corresponding  to chaotic or  disordered system. 
It turnes out, that the conductance can be expressed as $(1-c/2)N$ 
where the parameter $0<c<1$ measures the chaoticity of the system. 
\end{abstract}

In recent years  transport in a wide variety of  mesoscopic  systems has 
been  investigated. One of the  interesting questions  is the  behavior of  
the conductance of disordered identical blocks coupled in chain. This 
problem may arise whenever identical chaotic or disordered nanostructures are organized in a sequentially built 
structure. So far mostly chains with completely disordered/chaotic blocks have 
been studied\cite{Altshuler,Uzy}.  Our aim in this letter is to study  
a system in which  
chaoticity of the blocks can be tuned by varying a single parameter $c$  
ranging from zero (completely regular) to one (completely chaotic).   
Well known examples of such systems are provided by point-like impurities with  
variable strength placed in regular waveguides\cite{Haake,Seba}.   
We study here the low temperature conductance 
of a waveguide  containing finite number  
of periodically 
placed  identical  point-like  impurities. In this model  
the conductance depends on the strength and on the number of 
impurities.   
It will be shown that when passing from the regular toward   
the quantum-chaotic case by increasing the impurity strength 
the average conductance (in units of $2e^2/h$) behaves like   
$\sim \alpha N$, where  $\alpha=1-c/2$,  
interpolating between the conductance of the  regular (empty)  waveguide 
$N$  and  the universal conductance of  chaotic  
systems $\sim N/2 $ determined by Random Matrix Theory\cite{WDM} (RMT).  
The direct universal link between $c$, characterizing the chaoticity, and  
the average conductance is our main result.    
 
We consider an ideal 2D waveguide of width $W$ which is divided into $L$  
blocks of length $a$. We place an impurity with Dirac-delta potential  
$U(r)=\lambda\delta(r-r_0)$ 
in each block, where $r_0$ is its position within a block (see Fig. 1) 
and  
$\lambda$ is the strength of the potential. Inside the block the potential is  
assumed to be zero and the wave functions should fulfill Dirichlet boundary  
condition ($\psi=0$) on the walls of the  
waveguide. The system is adiabatically matched to 2D electron  
reservoirs on both ends.   
 
The conductance is given by the Landauer-B\"uttiker  
formula\cite{Landauer}                    
\begin{equation} 
G(E_F)=\frac{2e^2}{h}  T,   
\end{equation}    
where $T$ is the total transmission coefficient of the system at Fermi 
energy $E_F$. The total 
transmission can be expressed in terms of the partial transmission  
amplitudes $t_{nm}$ of open modes between the entrance and the 
exit of 
the waveguide 
\begin{equation}  
T=\sum_{n,m} | t_{nm}|^2 , 
\end{equation} 
where the transmission amplitudes can be calculated from the retarded  
Green-function of the system\cite{BarangerStone} 
\begin{equation}         
t_{nm}=2i(k_nk_m)^{1/2}\int dydy'\Phi_n(y)G^+(x,y|x',y')\Phi_m(y').    
\label{trans}   
\end{equation} 
Here $ \Phi_n(y)=\sqrt{\frac{2}{W}}\sin(\pi n y/W)$ is the transverse part  
of the wavefunction of the empty waveguide  
\begin{equation} 
\psi_n(x,y)=\Phi_n(y)e^{ik_nx},  
\end{equation} 
where $k_n=\sqrt{2mE_F-(\hbar\pi n/W)^2}$ for mode 
index $n$, 
$y$ is the transverse coordinate and longitudinal coordinates $x$ and $x'$  
lie anywhere on the entrance (left hand side in Fig. 1) and exit sides, 
respectively. The mode $n$ is open if $E_F>(\hbar\pi n/W)^2/2m$. 
 
The retarded Green-function can be calculated recursively 
\cite{Grosche}. Adding a new  
impurity to a system with $L$ impurities changes the  
Green-function  
the following way 
\begin{equation} 
G^+_{L+1}(x,y|x',y')=G^+_L(x,y|x',y')+ 
\tilde{\lambda} \frac{G^+_L(x,y|x_0,y_0) 
G^+_L(x_0,y_0|x',y')}{1-\tilde{\lambda} G^+_L(x_0,y_0|x_0,y_0)}, 
\end{equation}   
where the strength  $\tilde{\lambda}$ should be renormalized  
\begin{equation} 
\tilde{\lambda}=\frac{\lambda}{1+\lambda G^+_L(x_0,y_0|x_0,y_0,-E_F)}, 
\end{equation} 
in order to make the calculations with the delta potential well defined  
\cite{Grosche2}. 
The recursion starts with the retarded Green-function of the empty 
waveguide 
\begin{equation} 
G^+_0(x,y|x',y',E_F)=\sum_{n=1}^{n_{max}} 
\frac{\Phi_n(y)\Phi_n(y')}{2ik_n} e^{ik_n|x-x'|}, 
\end{equation} 
where $n_{max}\rightarrow\infty$. In numerical calculation $n_{max}$ should 
be much larger than the number of open modes $N=Int(k_F W/\pi\hbar)$.   
 
First we study the behavior of the conductance of the system with 
increasing number of blocks $L$ each containing one impurity of fixed 
strength $\lambda$.  
For the empty ideal waveguide  ($L=0$) the transmission as a function of 
the Fermi  wavelength $k_F=\sqrt{2mE_F}/\hbar$ is an ideal  staircase  
with steps $T=N$, where $N$ is the number of open modes.   
Adding one  impurity  ($L=1$) slightly  modifies the staircase\cite{Nakamura}  
and the transmission remains still close to $N$. 
By adding more impurities the staircase disappears and 
an irregular  structure emerges as it is shown in the upper 
part of Fig. 2 (solid line) for $L=10$. The transmission for 
$L\rightarrow\infty$ at a fixed Fermi wavelength decreases toward an  
asymptotic value which is the transmission of the infinitely 
long periodic system. 
 
The eigenmodes of the infinitely long periodic system are the Bloch functions 
\begin{equation} 
\psi_{q,n}(x,y)=e^{iqx}u_{q,n}(x,y), 
\end{equation}  
where $u_{q,n}(x+a,y)=u_{q,n}(x,y)$ and $-\pi/a<q<\pi/a$ with 
eigenenergies $\epsilon_n(q)$ forming a band structure. 
In this case the transmission is given 
by the number of Bloch modes $N_B$ propagating in 
positive $x$ direction at Fermi energy $E_F$, which can be determined 
by counting the number of solutions $(n_i,q_i), 
i=1,...,N_B$ for $q_i>0$ of  
\begin{equation} 
\epsilon_{n_i}(q_i)=E_F. 
\end{equation} 
In the lower part of Fig. 2 the band structure for a given 
value of $\lambda$ is shown for $q>0$. Here one can count the 
number of intersections of bands with $k_F$ yielding $N_B$ and 
the transmission $T=N_B$ of the infinite system is 
plotted on the upper part of Fig. 2 (dashed line).  
We can see that the transmission, even for $L=10$, approaches  
integer values corresponding to the transmission of the infinite 
system. The transmission drops to zero when $k_F$ is in 
a gap of the band structure and it jumps whenever 
$k_F$ is on the edge of a band. 
 
Next we study the  $L$ dependence of the average transmission $<T>_N$  
corresponding to a fixed number of open modes $N$, {\em i.e.} for  
Fermi wavelengths from $k_F=N$ to $k_F=N+1$ (hereafter measured in 
units of $\pi\hbar/W$), 
\begin{equation} 
<T>_N=\int_{N}^{N+1} T(k_F)dk_F. 
\end{equation}    
In Fig. 3 $<T>_{10}$ is shown as a function of the number of blocks $L$ 
for different values of $\lambda$.  
The average transmission as the function of $L$ decreases 
exponentially and can be well approximated with the expression   
\begin{equation} 
<T>_N = (N-T_\infty(\lambda,N))e^{-L/\xi(\lambda,N)}+T_\infty(\lambda,N), 
\end{equation} 
where $\xi$ characterizes the rate of decay and $T_\infty$ is 
the number of modes 
remaining open for $L\rightarrow\infty$.  
We can interpret $\xi$ as a partial localization length in the 
sense that the conductance decays exponentially with the  
size of the system, however it does not decay to zero,  
like in case of strong-localization, due to the periodicity of  
the system. 
 
In Fig. 4 we plotted the asymptotic transmission $T_\infty$ as   
a function of the number of open modes $N$ for different 
values of $\lambda$. The transmission $T_\infty$ seems to scale 
approximately linearly with $N$. For large $\lambda$ it is in  
good agreement 
with the prediction of the Circular Orthogonal Ensemble (COE) 
of RMT for large N 
\begin{equation} 
T_{COE}\approx \frac{N}{2}-\frac{1}{4}, 
\label{COE} 
\end{equation} 
worked out for chaotic cavities with time reversal  
symmetry in Ref.\cite{Baranger2}. 
For other values of $\lambda$ one can fit  
\begin{equation} 
T_\infty(\lambda,N) = \alpha(\lambda) N + \beta(\lambda).\label{fit} 
\end{equation} 
 
In Fig. 5 the fitted parameters $\alpha(\lambda)$ and $\beta(\lambda)$ 
are shown. One can see that $\beta$ is approximately independent 
of $\lambda$ and takes the value $-1/4$, the weak localization 
correction, predicted by RMT. This means, that this weak localization 
correction sets in if back scattering is possible 
($\lambda\neq 0$), however it does not depend on the details of its 
mechanism. 
 
The parameter $\alpha$ decreases 
from $\alpha\approx 1$ to $\alpha\approx 1/2$ with increasing $\lambda$. 
The $\lambda$ dependence of $\alpha$ can be explained by making 
a simple assumption about the relation between the conductance  
and the chaoticity of the system as we are going to show next. 
The Hamiltonian in systems which are neither completely regular nor 
chaotic can be divided into a regular $H_R$ 
and a chaotic $H_C$ part 
\begin{equation} 
H=H_R+H_{C}. 
\end{equation} 
In our case $H_R$ is the Hamiltonian of a block of the empty waveguide  
and $H_C$ is associated with the delta impurity in it. The regular part 
can be modeled with a diagonal matrix with random elements and mean 
level spacing  
$\Delta$. The chaotic part can be modeled with a random Gaussian matrix 
drawn from the suitable orthogonal ensemble (GOE, GUE etc.) 
reflecting the symmetry of the system. The variance of the 
elements of the random matrix should coincide with the variance 
of the matrix elements of the chaotic part of the Hamiltonian {\em i.e.}
$\sigma^2=\overline{|<i|H_C|j>|^2}$. The energy level statistics of 
such systems is transitional between Poissonian and tpure RMT. 
The chaoticity of the system, reflected in the level statistics,  
is determined by the transition parameter $\eta=\pi\sigma/\Delta$
whose universality was shown in many applications\cite{appl}.
The appropriate measure of the chaoticity of a system is the 
rescaled parameter defined in Ref.\cite{Guhr} 
\begin{equation} 
c=\frac{\eta}{\sqrt{1+\eta^2}} 
\end{equation} 
ranging from $0$ to $1$. The value $c=0$ corresponds to 
a completely regular system (the empty waveguide) and 
for $c=1$ the system is completely chaotic ($\lambda 
\rightarrow \infty$) in a quantum-chaotic sense. 
Analogously, in our numerical calculation ($W=a=1$), the chaoticity  
parameter $c$ can be defined as 
\begin{equation} 
c(\lambda)=\frac{\lambda/4}{\sqrt{1+(\lambda/4)^2}}. 
\label{oc} 
\end{equation} 
Using semiclassical considerations for large $N$  
we can make connection between the chaoticity parameter $c$ and 
the parameter $\alpha$ characterizing $T_\infty$ for $N\rightarrow\infty$. 
Semiclassically a part of the trajectories is bouncing between the 
walls only, while the rest is scattered by the impurities irregularly. 
Accordingly, the transmission can be approximately partitioned 
into a regular and a chaotic part. We can assume that the effective 
number of chaotic modes is $N\cdot c$ while the number of regular modes 
is $N\cdot (1-c)$. For  regular modes the propagation is ideal, free from 
reflection and their contribution to the transmission is 
$N\cdot (1-c)$. On the other hand, according to the RMT result 
(\ref{COE}), only half of the chaotic modes 
contribute to the transmission and we get 
\begin{equation} 
T_\infty\approx N\cdot (1-c)+\frac{1}{2}N \cdot c , 
\label{formula} 
\end{equation} 
yielding  
\begin{equation} 
\alpha=1-c/2. 
\label{alp} 
\end{equation} 
 
In Fig. 5 we compared the numerically fitted values of $\alpha(\lambda)$ 
(see Eq. (\ref{fit})) to $1-c(\lambda)/2$ (see Eq. (\ref{oc})) and we 
have found excellent agreement in the whole range $\lambda=1 - 500$ 
supporting our new formula (\ref{alp}). 
 
We conjecture that our new formula (\ref{alp}) is valid for the  
transmission of a much broader class of systems. 
It is expected that it works for the transmission through a system,  
attached to leads, 
whose level statistics (without leads) is transitional\cite{Guhr}  
in between Poissonian and RMT  and can be characterized with $0<c<1$, 
since the assumption made above are valid in this case too. 
This makes it possible to determine the chaoticity of a system 
from the relation 
\begin{equation} 
c=2\cdot (1-\alpha), 
\end{equation} 
by estimating $\alpha$ from the conductance. 
We hope that further numerical and possible experimental studies 
will support the results presented here. 
 
This work has been supported by the Hungarian Science Foundation OTKA  
(F019266/F17166/T17493) the Hungarian 
Ministry of Culture and Education FKFP 0159/1997 and the 
Hungarian-Israeli R\&D collaboration project (OMFB-ISR 96/6). J. Cs. thanks
the financial support of The Royal Society of London.

\pagebreak
\begin{figure} 
\centerline{\strut\psfig{figure=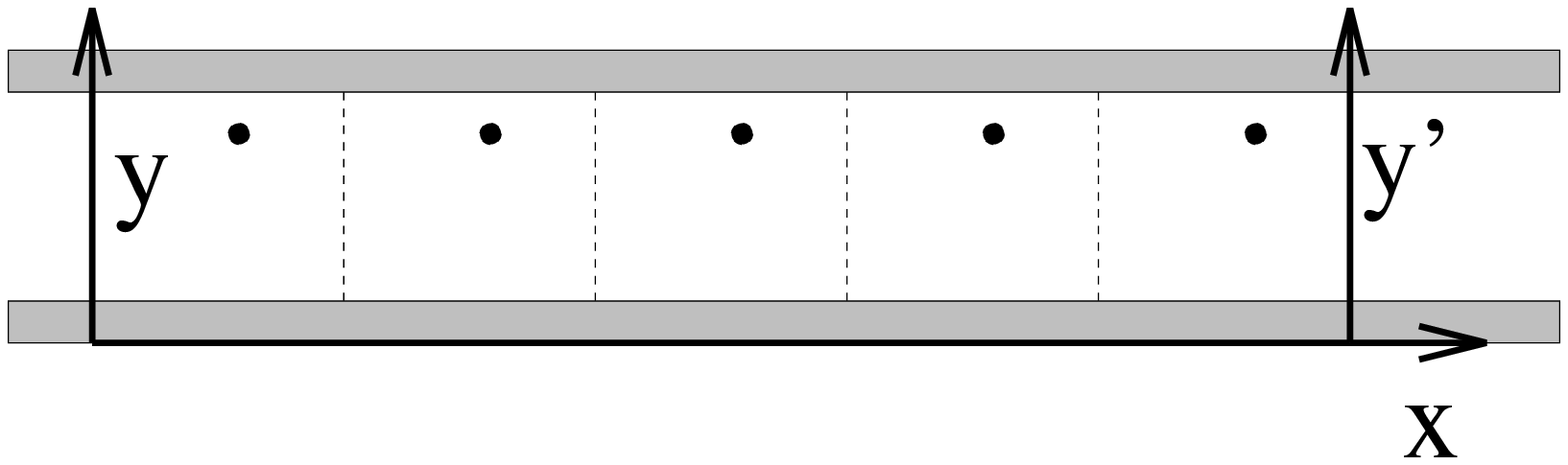,width=70mm}} 
\caption{Schematic picture of the waveguide containing periodically 
placed impurities. The distance between the impurities is $a$ and 
the width of the guide is $W$. The entrance is on the left hand side. 
In the numerical calculations we choose $W=a=1$ and the position 
of the impurities was $x_0=0.65$ and $y_0=0.84$ with respect 
to the left lower corner of the unit cell.} 
\label{fig1}  
\end{figure} 
 
\pagebreak
 
\begin{figure}
\centerline{\strut\psfig{figure=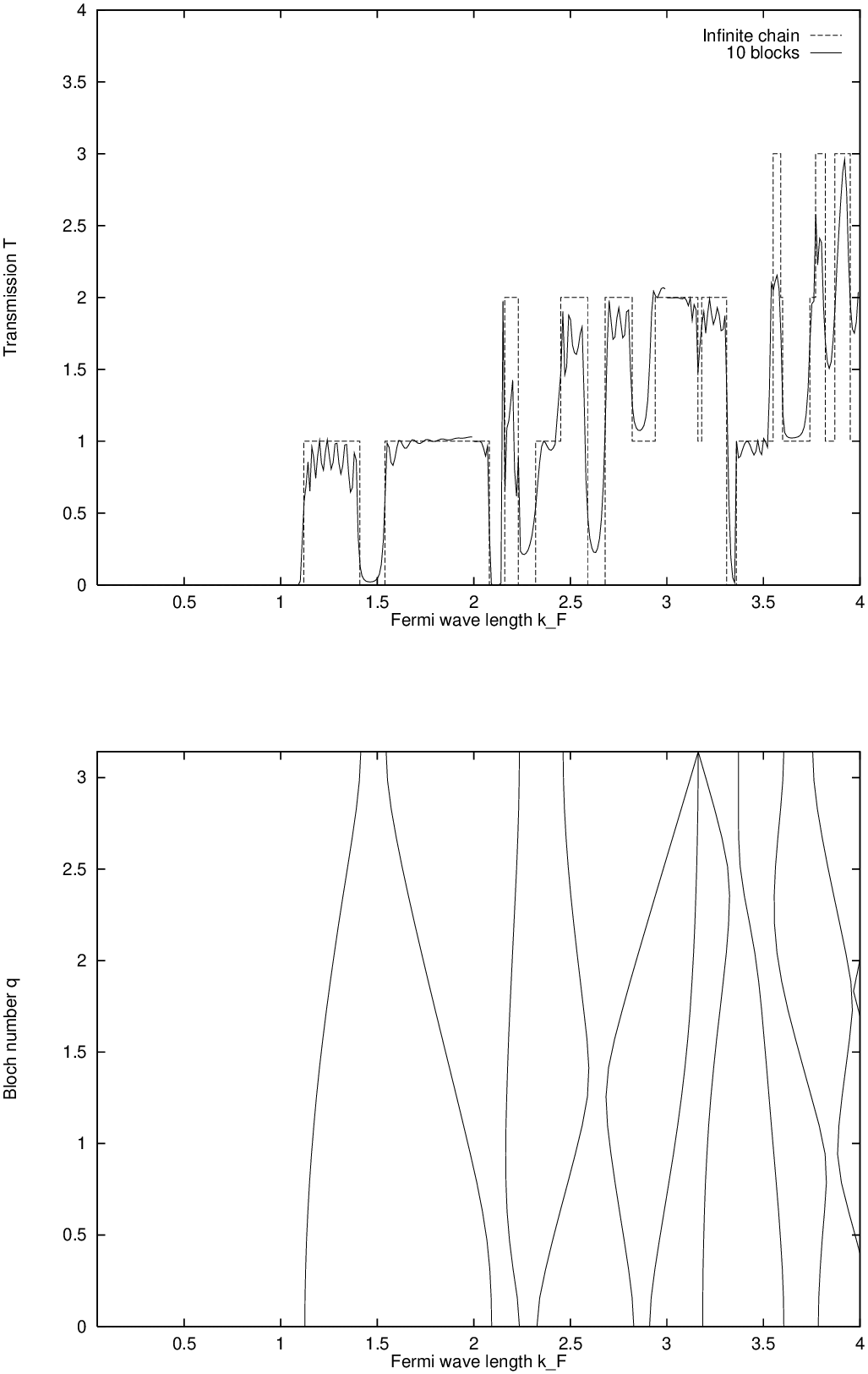,width=130mm}}     
\caption{In the upper part the transmission vs. the Fermi wavelength 
in units of $\hbar\pi/W$ is plotted for a system with $L=10$ blocks 
(solid line) and for an infinitely long system $L=\infty$  
(dashed line). $\lambda=5$ in both cases. 
The rapid fluctuations of the conductance are on the scale $\Delta k\sim 
1/La$ caused by the  multiple scattering between the two ends of the device. 
In the lower part the band structure of the infinitely 
long system is plotted for the same parameter $\lambda$ as above. 
The vertical axis is the Bloch number $q$ and  
$ k_n (q)=(2m\epsilon_n (q))^{1/2}   /\hbar $  
has been plotted in units of  
$\hbar\pi/W$. This representation makes it possible to get the 
transmission of the infinitely long system in the upper part from 
the bands of the lower part graphically. }  
\label{fig2}  
\end{figure} 
 
\pagebreak
 
\begin{figure}
\centerline{\strut\psfig{figure=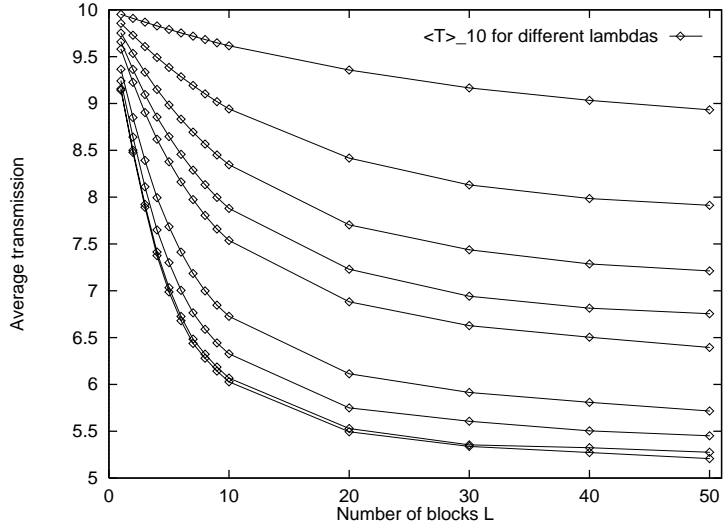,width=70mm}}     
\caption{The average transmission for $N=10$ open modes as 
a function of the number of blocks $L$ at different 
values of impurity strength $\lambda$ and for $W=a=1$.  
$<T>_{10}$ decreases from $10$ (the transmission of the 
empty waveguide) exponentially as a  
function of $L$ for all $\lambda$. Around $L\approx 50$ it 
reaches an asymptotic value depending on $\lambda$. For large 
$\lambda$ (chaotic case) this asymptotic value is close to $N/2=5$ 
predicted by Random Matrix Theory.}  
\label{fig3}  
\end{figure} 
 
\begin{figure} 
\centerline{\strut\psfig{figure=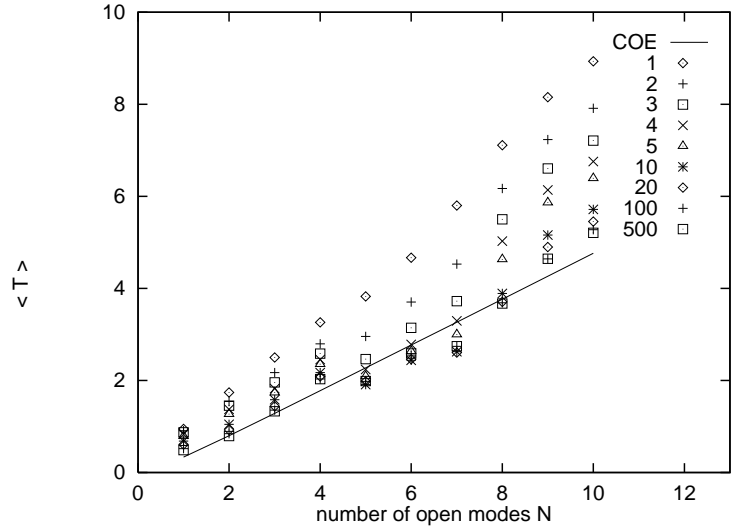,width=70mm}}    
\caption{The asymptotic transmission $T_{\infty}$ as a function 
of the number of open modes $N$ of the empty waveguide for  
different values of $\lambda$ in the range of $1-500$. 
The solid line represents the COE result $T=N^2/(2N+1)$. }  
\label{fig4}  
\end{figure} 
\pagebreak

\begin{figure} 
\centerline{\strut\psfig{figure=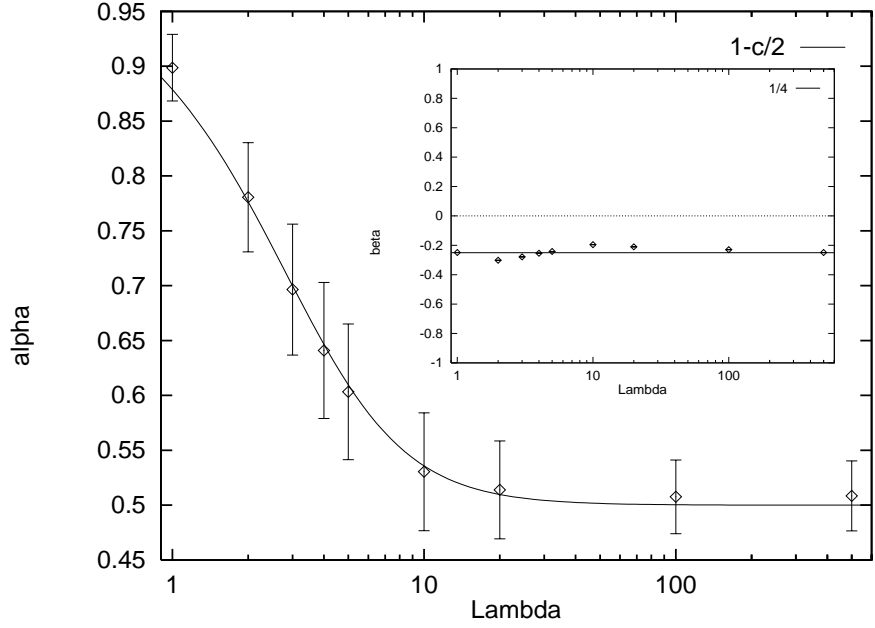,width=120mm}}    
\caption{ The parameters $\alpha$ and $\beta$ (see Eq. (\ref{fit})) 
of the linear fit to the data of Fig. 4 for different values of $\lambda$ in 
the range of $1-500$. {\bf a,} parameter $\alpha$ vs. $\lambda$ 
(diamonds) and our prediction $\alpha=1-c(\lambda)/2$ (see. Eq. 
(\ref{formula})). The errorbars represent the error of the linear fit.  
{\bf b,} parameter $\beta$ versus $\lambda$ 
(diamonds) and the weak localization correction $-1/4$ (solid line).}  
\label{fig5}  
\end{figure}

\end{document}